\documentstyle [11pt,epsf]{article}
\begin{document}
%
%
\newcommand{\Abs}[1]{|#1|}
\newcommand{\EqRef}[1]{(\ref{eqn:#1})}
\newcommand{\FigRef}[1]{fig.~\ref{fig:#1}}
\newcommand{\Abstract}[1]{\small
   \begin{quote}
      \noindent
      {\bf Abstract - }{#1}
   \end{quote}
    }
\newcommand{\FigCap}[2]{
\ \\
   \noindent
   Figure~#1:~#2
\\
   }
%
%
%
%
\title{On the decay of correlations in Sinai billiards with infinite horizon}
\author{
Per Dahlqvist \\
Mechanics Department \\
Royal Institute of Technology\\ 
S-100 44 Stockholm, Sweden\\[0.5cm]
and\\[0.5cm]
Roberto Artuso\\
Istituto di Scienze Fisiche, Chimiche e Matematiche\\
Universit\'a degli Studi di Milano\\
Via Lucini, 3, I-22100 Como, Italy\\
}
\date{}
\maketitle

\ \\ \ \\

%
\Abstract{We compute the decay of the autocorrelation 
function of the observable
$|v_x|$ in the Sinai billiard and of the observable $v_x$ in the associated
Lorentz gas with an approximation due to Baladi, Eckmann and Ruelle. 
We consider the standard
configuration
where the disks is centered inside a unit square.
The
asymptotic decay is found to be $C(t) \sim c(R)/t$. An explicit expression is
given for the prefactor $c(R)$ as a function of the radius of the scatterer.
For the small scatterer case we also
present expressions for the preasymptotic regime.
Our findings are supported by numerical computations.

%
}

\ \\
\section{Introduction}

In this paper we are going to present analytical and
numerical evidence that typical correlation functions in a Sinai
billiard with infinite horizon asymptotically decay as $C(t) \sim c(R)/t$.
On the theoretical side there has been some heuristic
argument \cite{Frid,ZGN} 
and rigorous bounds \cite{Bum}
suggesting this type of behavior. 
We will strengthen the argument for the suggested type of decay  
and derive an expression for the constant $c(R)$ as a function of disk
radius $R$. 
On the numerical side we investigate correlation functions via Monte Carlo 
simulations: a remarkably good agreement is found with theoretical 
expressions, which accurately predict the observed long-time behaviour, and
also account for the main features of the preasymptotic region.
The $1/t$ decay of the velocity autocorrelation function was, to our 
knowledge, first observed numerically in \cite{Frid}, and confirmed, 
in a somehow different approach, focusing on power spectrum behaviour, 
in \cite{ZGN}. More recently the asymptotic algebraic decay was 
detected in \cite{Art2}, while only the preasymptotic regime was 
investigated in the simulations described in \cite{Gala}.

We will focus on two particular correlation functions. First the
autocorrelation function of the observable $|v_x|$ in the Sinai billiard.
The modulus is taken to prevent the observable to change sign by the frequent
bounces on the square walls which would make our calculations more difficult.
Secondly we study the autocorrelation of the observable $v_x$ in the associated
Lorentz gas, the difference from the former case is that the sign of the
observable may now be changed by bounces on the disk(s).

The theoretical calculations will be carried out 
following an idea originally due to
Baladi, Eckmann and Ruelle \cite{BER,PDreson,PDsin,PDLA,PDlyap}.
This approximation (called the BER approxmation) for
intermittent systems divide the temporal evolution
of the systems into {\em intervals}, each interval consist of one
laminar and one chaotic interval.
The approximation assumes the absence of correlations between
disjoint intervals. In particular the lengths $\Delta_i$ are assumed 
mutually independend
and one can define a probability distribtuion $p(\Delta)$, a function that
encodes a lot of information about the system.
For the observables considered in this paper the application of this
approximation is almost trivial.

\section{Application to the Sinai billiard}

We will consider the version of the
Sinai billiard \cite{Sin} consisting of a unit square with a 
scattering disk, having radius $R$;
$0<2R \leq 1$, centered on its midpoint.
This is a system with infinite horizon, for any finite radius $R$ there is
a finite number of corridors, though which a particle may go without ever
bounce onto a disk (these never-touching trajectories
constitute a set of measure zero in phase space).

We let the point particle have unit velocity.
The trajectory of a particle in
the Sinai billiard consists of laminar intervals,
(bouncing between the straight sections) interrupted by scatterings
off the central disk. The variable 
$\Delta$ introduced above is simply the length
of the trajectory {\em segment} between two
disk bounces.

\subsection{The distribution of reccurence times}

We introduce the disk as surface of section with phase space coordinates
$x_s$. We define $\Delta_s(x_s)$ as the traveling distance to the 
next disk bounce.
The distribution $p(\Delta)$ is the distribution of recurrence times
to the surface of section
\begin{equation}
p(\Delta)=\frac{1}{V_s} \int dx_s \; \delta (\Delta-\Delta_s(x_s))  \ \ ,
\label{eqn:pDelta}
\end{equation}
where $V_s=\int dx_s=4\pi R$ using Birkhoff coordinates.

The relevant time scale is set by the average
$<\Delta>\equiv \int \Delta p(\Delta) d\Delta$ which can be computed exactly as
follows.
When we integrate over phase space (or rather the
energy surface) it is convenient to split up the
phase space element according to $dx=dx_s d\tau$ where $\tau$ is is the
coordinate along the the trajectory. The total phase space volume may now be
expressed as 
\begin{equation}
V=\int dx= \int dx_s \int_0^{\Delta_s(x_s)}d\tau=
\int \Delta_s(x_s) dx_s=V_s <\Delta>=4\pi R <\Delta>  \  \ .
\end{equation}
On the other hand we have
\begin{equation}
V=\int dx\; dy\; dv_x dv_y \delta(1-\sqrt{v_x^2+v_y^2})=
(1-\pi R^2)2\pi  \  \  ,
\end{equation}
leading to the following expression for the expectation value of $\Delta$
\begin{equation}
<\Delta>=\frac{1}{2R}-\frac{\pi R}{2}  \  \  . \label{eqn:Deltaexp}
\end{equation}

In ref. \cite{PDsmall} we derived the following expression for
$p(\Delta)$ in the the small $R$ limit
\begin{equation}
p_{R\rightarrow 0}(\Delta)= \left\{ \begin{array}{ll}
\frac{12R}{\pi^2} & \xi<1\\
\frac{6R}{\pi^2 \xi^2}(2\xi+\xi(4-3\xi)\log(\xi)+
4(\xi-1)^2 \log(\xi-1)-(2-\xi)^2\log|2-\xi|)&
\xi>1 \end{array} \right.   \  \  ,  \label{eqn:smallR}
\end{equation}
where
$\xi=\Delta/<\Delta>$.
Some smearing in $\xi$ is required to make the limit well defined
\cite{PDsmall}.

For finite R there is of course no such simple formula, 
but the tail ($\Delta \gg <\Delta>$) exhibit the asymptotic law
\begin{equation}
p(\Delta) \sim  \frac{4\sigma(R)}{\pi\Delta^3} \  \  , \label{eqn:pDeltaas}
\end{equation}
where $\sigma(R)$ is the sum
\begin{equation}
\sigma(R)=  \sum_{{\bf q} \in S}
(2qR+\frac{1}{2qR}-2)   \  \  .  \label{eqn:sigma}
\end{equation}
where
\begin{equation}
S=\{ {\bf q}=(n_x,n_y) | n_y>0; \; n_x \geq 0; \;  (n_x,n_y)=1; \; 
q\equiv\sqrt{n_x^2+n_y^2}<1/2R  \}
\end{equation}
is the set of corridors.

The leading asymptotic behaviour of the function $\sigma(R)$ is in ref.
\cite{Bleh} found to be $\sigma(R) \sim 1/(4\pi R^2)$ and in this limit we get
\begin{equation}
p(\Delta) \sim  \frac{1}{\pi^2 R^2 \Delta^3} \; \; \; \; R \rightarrow 0
\  \  .  \label{eqn:pDeltaasas}
\end{equation}

\subsection{Correlation functions in the BER approixation}

 We will restrict ourselves
to observables $A$ 
that are changed only by bounces on the disk.
The autocorrelation function is e.g. obtained as the time average
\begin{equation}
C_{AA}(t)=\langle A(t_0+t)A(t_0) \rangle_{t_0}-\langle A \rangle^2  \ \ .
\end{equation}
The BER approximation assumes that there is no correlations
if $t$ and $t+t_0$ belongs to different intervals, that is
$\langle A(t_0+t)A(t_0) \rangle=\langle A^2 \rangle$ if there is no bounce
between $t$ and $t+t_0$ and\linebreak
$\langle A(t_0+t)A(t_0) \rangle=\langle A \rangle^2$
otherwise.

Next let $P_0(t)$ denote
the probability that the trajectory has {\em not} hit
the disk between $t$ and $t+t_0$.
We can now write the correlation function in terms of conditional probabilities
\begin{equation}
C_{AA}(t)=P_0(t)\langle A^2 \rangle+(1-P_0(t))\langle A \rangle^2
-\langle A \rangle^2=
P_0(t)V(A)   \ \ , \label{eqn:nice}
\end{equation}
where $V(A)$ is the variance of $A$.

The function $P_0(t)$ may be expressed in terms of $p(\Delta)$ 
\begin{equation}
P_0(t)=\frac{1}{\langle \Delta \rangle}
\int_t^\infty \{ \int_u^\infty p(\Delta)d\Delta \} du  \ \ .
\label{eqn:p0}
\end{equation}
The $1/\Delta^3$ decay of $p(\Delta)$ thus implies a $1/t$ decay of the
correlation function.
From eq \EqRef{pDeltaas} we compute the tail of $P_0(t)$
\begin{equation}
P_0(t) \sim \frac{2\sigma (R)}{<\Delta>\pi} \frac{1}{t}  \   \   .
 \label{eqn:p0as}
\end{equation}

In the $R \rightarrow 0$ limit we get from eqs. \EqRef{p0},
\EqRef{pDeltaasas} and \EqRef{Deltaexp}
\begin{equation}
P_0(t) \sim \frac{1}{\pi^2 R} \frac{1}{t}\; \; \; \; 
R \rightarrow 0  \  \  .  \label{eqn:p0asas}
\end{equation}

For sufficiently small
$R$ one can use eq. \EqRef{smallR} to obtain an expression for the entire
$P_0(t)$ (the resulting expression
is not particularly nice so we do not display it).
For other radii $R$ one can use
a numerically obtained $p(\Delta)$ to compute $P_0(t)$.

First we will study the observable $A=|v_x|$. Since 
$\langle |v_x|^2 \rangle=1/2$ and $\langle |v_x| \rangle=2/\pi$ we get
\begin{equation}
C_{AA}(t)=
P_0(t)(\frac{1}{2}-\frac{4}{\pi^2})   \ \ , \label{eqn:CAA}
\end{equation}

\subsection{Correlations in the Lorentz gas}

A closely related problem is the decay of correlations in the associated
Lorentz gas obtained by unfolding the bounded billiard into an infinite lattice
of disks, cf. fig 1.
If we kept considering the observable $A=|v_x|$ we 
would just recover the previous result.
Let us instead consider the observable $B=v_x$. The difference is
that the bounces on disks might now change the sign of $B$. The expression
\EqRef{nice} may again be used, with the difference that  
$\langle B \rangle =0$
so we get
\begin{equation}
C_{BB}(t)=P_0(t)\cdot 
\left( \langle B^2 \rangle-\langle B \rangle^2\right)=P_0(t)/2  \  \  . 
\label{eqn:CBB} 
\end{equation}

It is well-known that this correlation function is related to the diffusion
coefficient
\begin{equation}
D= \lim_{t \rightarrow \infty}\frac{1}{ft}<(\bar{x}(t)-\bar{x}(0))^2>   \  \  ,
\end{equation}
via the Einstein-Green-Kubo formula \cite{Bleh}
\begin{equation}
D=\lim_{t \rightarrow \infty} \frac{2}{f} \int_0^t <\bar{v}(0)\cdot\bar{v}(s)>ds
=\lim_{t \rightarrow \infty} 2 \int_0^t <v_x(0)\cdot v_x(s)>ds
\end{equation}
where we have made use of the symmetry of our system
$<v_x(0)\cdot v_x(s)>=<v_y(0)\cdot v_y(s)>$.\\
$f$ is the number of degrees of
freedom.

Inserting eq. \EqRef{CBB} into this formula we obtain the diverging
diffusion constant
\begin{equation}
D(t)= \frac{2 \sigma(R)}{<\Delta >\pi}\log t  \  \  ,
\end{equation}
in agreement with refs. \cite{PDlyap,Bleh}

\subsection{Comparison with numerical data}

The numerical experiments are carried out through Monte Carlo phase 
integration (based on a subtractive random generator), using $10^7$ initial
conditions. Data become noisy when correlations decrease to  the same order of
magnitude as the integration error (which is inversely  proportional to the
square root of the number of initial conditions). It has been pointed out (see
for example \cite{Ru} and \cite{Gala}) that in principle one might worry about
errors due to exponential propagation of  errors (due to positivity of the
Lyapunov exponent and finite precision  arithmetics): though our time sequences
extend beyond the limit suggested  by former arguments, the onset of power-law
(asymptotic) decay is well  within the allowed time scale (where the data are
insensitive to  statistical errors, which in turn dominate over the afore
mentioned  errors). We also remark that for hyperbolic billiards
Lyapunov-induced  errors did not seem to lead to detectable effects, see
\cite{Art,Art2}.

In fig. 2 and 3 we show typical results of our numerical experiments: the 
numerical correlation function indeed exhibits a $1/t$ decay for long 
times. In the 
same figure we also plot theoretical curves obtained from eq. 
\EqRef{CAA}. The good agreement, even along the preasymptotic region, 
is suggestive of the applicability of BER approximation for such a system. 
From fig. 2 and 3 one can observe that BER expressions, like
\EqRef{CAA} are not able to reproduce preasymptotic oscillations in
the correlation functions, while being quite close to their
envelope, even for moderate time. Such oscillations are smaller 
and smaller in the $R\rightarrow 0$ limit.
We comment on the deviation for large $t$ in fig 3 in section 3.

In the small radius limit, eq. \EqRef{smallR} suggests a scaling 
form of the correlation function $C_{AA,R}(t)=g(Rt)$ where
$g$ is independent of the value of $R$: in fig. 4 we show how the scaling 
is approximately correct for small times, and indeed is well confirmed in 
the asymptotic region.

Finally a comparison between correlation functions for A and B observables
is presented in fig. 5.

\section{Concluding remarks}

There are two obvious ways of refining
the present computations without abandoning the basic idea of the BER
approximation.

There is a tacit assumption about isotropy behind the factorization
$(1-P_0(t))\langle A \rangle^2$ in eq. \EqRef{nice}.
It is true that
the average of $|v_x|$ is exactly $ =2/\pi$
but only provided the average is taken over the
entire phase space.
But suppose we restrict the phase space integral to points lying on trajectories
penetrating deep into the corridors then it is no longer exactly true that
$\langle |v_x| \rangle =2/\pi$ (except in the small $R$ limit).
This will lead to a small correction of the prefactor $c(R)$ of the asymptotic
limit $C(t)\sim c(R)/t$ for observable A but not for observable B.
Preliminary results indicate that this effect account for
the deviation in the large $t$ limit in fig. 3 and fig. 5.

Secondly, if we have a very long segment into one corridor, than there
is an enhanced probability that the next segment goes into the same
corridor \cite{Bum}. Consecutives segments in the same corridor should be
considered as belonging to the same interval as they are associated with nearly
the same value of the observable.
This effect will provide a small
correction to $p(\Delta)$.
We will return to these problems in future work.
\vspace{0.5cm}

PD was supported by the Swedish Natural Science
Research Council (NFR) under contract no. F-FU 06420-303.
RA was partially supported by INFN, sezione di Milano and INFM, 
Unita' di Milano.

\newcommand{\PR}[1]{{Phys.\ Rep.}\/ {\bf #1}}
\newcommand{\PRL}[1]{{Phys.\ Rev.\ Lett.}\/ {\bf #1}}
\newcommand{\PRA}[1]{{Phys.\ Rev.\ A}\/ {\bf #1}}
\newcommand{\PRD}[1]{{Phys.\ Rev.\ D}\/ {\bf #1}}
\newcommand{\PRE}[1]{{Phys.\ Rev.\ E}\/ {\bf #1}}
\newcommand{\JPA}[1]{{J.\ Phys.\ A}\/ {\bf #1}}
\newcommand{\JPB}[1]{{J.\ Phys.\ B}\/ {\bf #1}}
\newcommand{\JCP}[1]{{J.\ Chem.\ Phys.}\/ {\bf #1}}
\newcommand{\JPC}[1]{{J.\ Phys.\ Chem.}\/ {\bf #1}}
\newcommand{\JMP}[1]{{J.\ Math.\ Phys.}\/ {\bf #1}}
\newcommand{\JSP}[1]{{J.\ Stat.\ Phys.}\/ {\bf #1}}
\newcommand{\AP}[1]{{Ann.\ Phys.}\/ {\bf #1}}
\newcommand{\PLB}[1]{{Phys.\ Lett.\ B}\/ {\bf #1}}
\newcommand{\PLA}[1]{{Phys.\ Lett.\ A}\/ {\bf #1}}
\newcommand{\PD}[1]{{Physica D}\/ {\bf #1}}
\newcommand{\NPB}[1]{{Nucl.\ Phys.\ B}\/ {\bf #1}}
\newcommand{\INCB}[1]{{Il Nuov.\ Cim.\ B}\/ {\bf #1}}
\newcommand{\JETP}[1]{{Sov.\ Phys.\ JETP}\/ {\bf #1}}
\newcommand{\JETPL}[1]{{JETP Lett.\ }\/ {\bf #1}}
\newcommand{\RMS}[1]{{Russ.\ Math.\ Surv.}\/ {\bf #1}}
\newcommand{\USSR}[1]{{Math.\ USSR.\ Sb.}\/ {\bf #1}}
\newcommand{\PST}[1]{{Phys.\ Scripta T}\/ {\bf #1}}
\newcommand{\CM}[1]{{Cont.\ Math.}\/ {\bf #1}}
\newcommand{\JMPA}[1]{{J.\ Math.\ Pure Appl.}\/ {\bf #1}}
\newcommand{\CMP}[1]{{Comm.\ Math.\ Phys.}\/ {\bf #1}}
\newcommand{\PRS}[1]{{Proc.\ R.\ Soc. Lond.\ A}\/ {\bf #1}}
%


\newpage

\section*{Figure captions}

\FigCap{1} {The Lorentz gas, obtained by unfolding the Sinai billiard over
the plane.}

\FigCap{2}{Experimental correlation function (full line), for $R=0.318$.
The dashed dotted line
represents the BER approximation \EqRef{CAA}, using a numerical $p(\Delta)$.}

\FigCap{3}{The full line represents the numerical correlation function
for $R=0.106$,
the dotted line is obtained from \EqRef{CAA}, using the theoretical small $R$
limit of $p(\Delta)$ 
\EqRef{smallR}, while the dashed line has been obtained by using the
asymptotic
expression \EqRef{p0as}.}

\FigCap{4}{Correlation functions for different $R$ values, plotted as
functions of $Rt$. The dashed line corresponds to $R=0.053$, the dashed
dotted line to $R=0.106$ and the full line to $R=0.212$.}

\FigCap{5} {Numerical correlation function for observable B (upper full
line) and A (lower full line) for the Lorentz gas (R=0.159). The dashed lines
refer to the asymptotic expression \EqRef{p0as} fed into equations \EqRef{CAA}
and \EqRef{CBB}.}

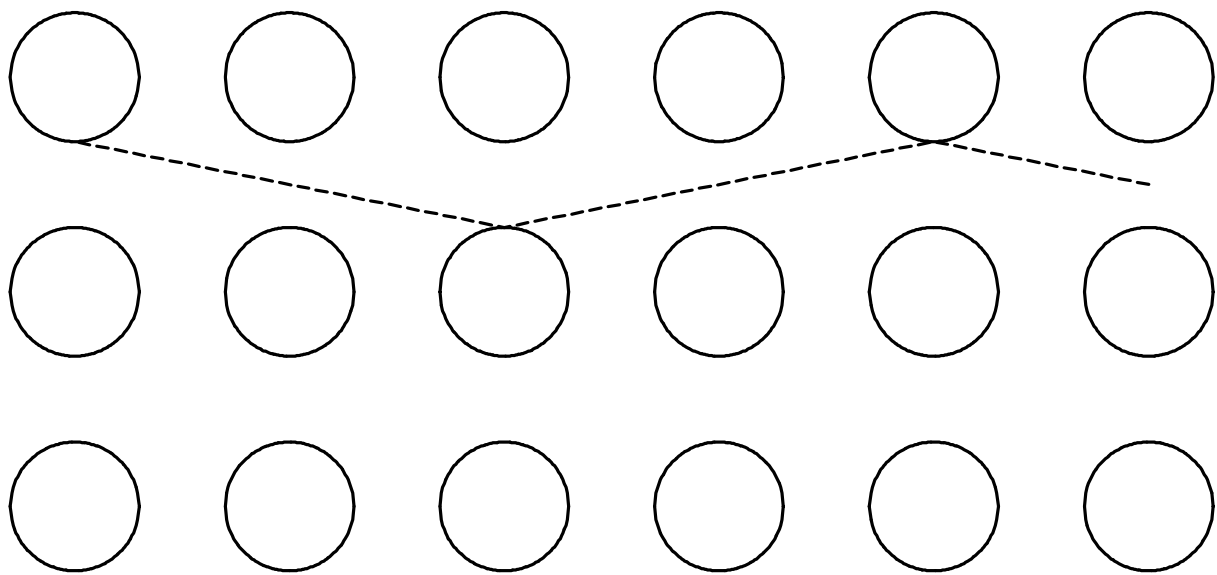
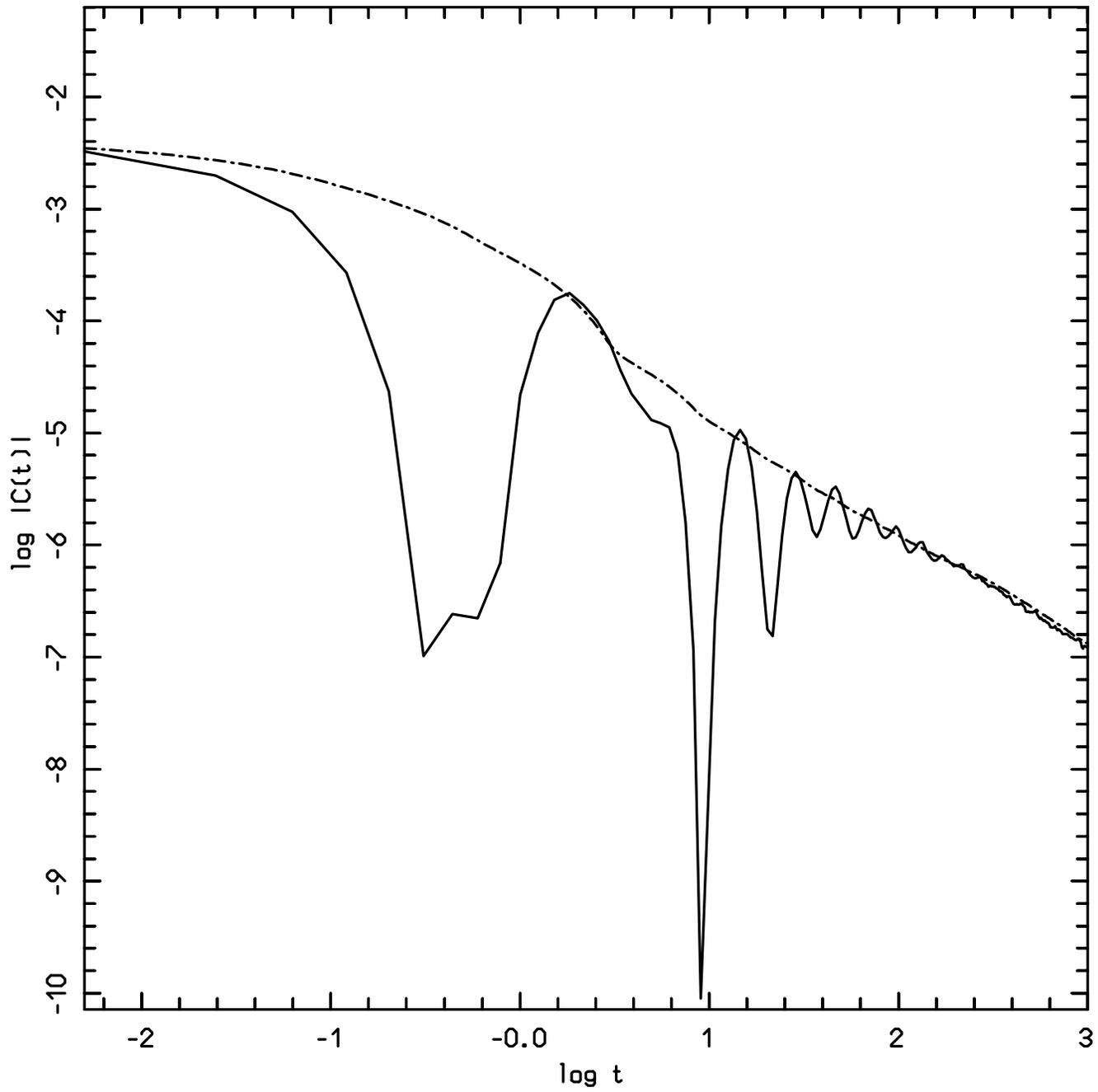
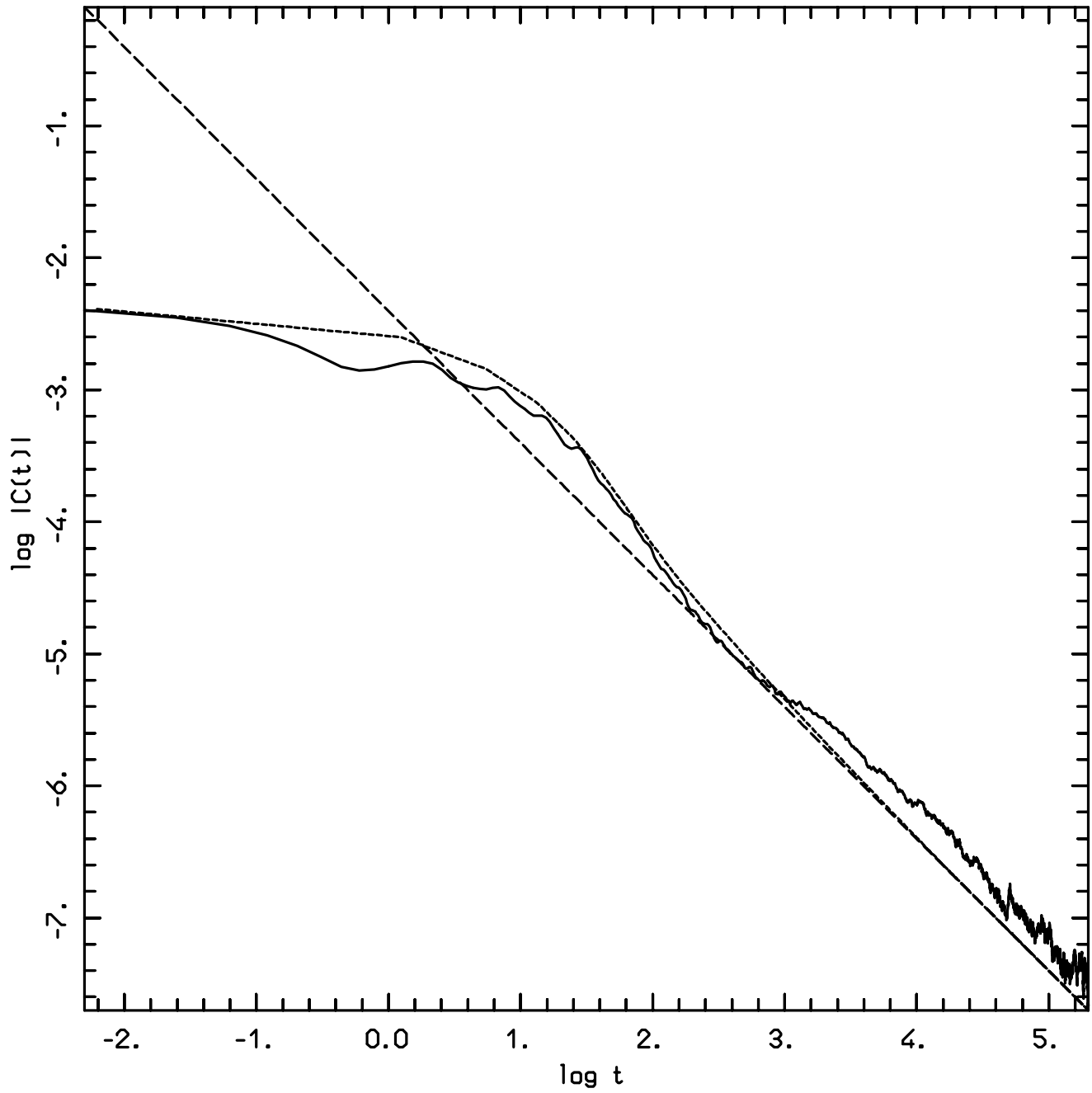
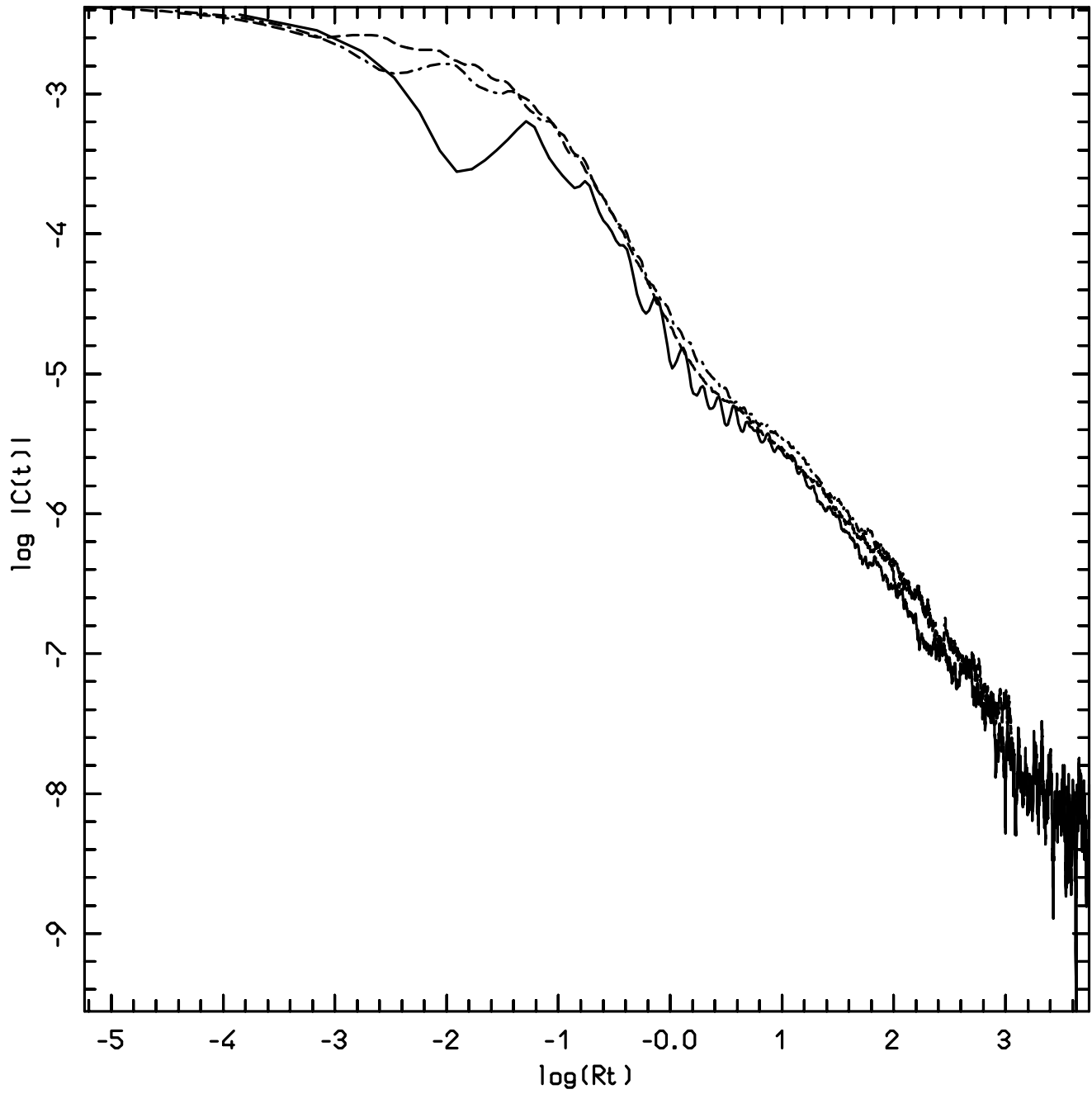
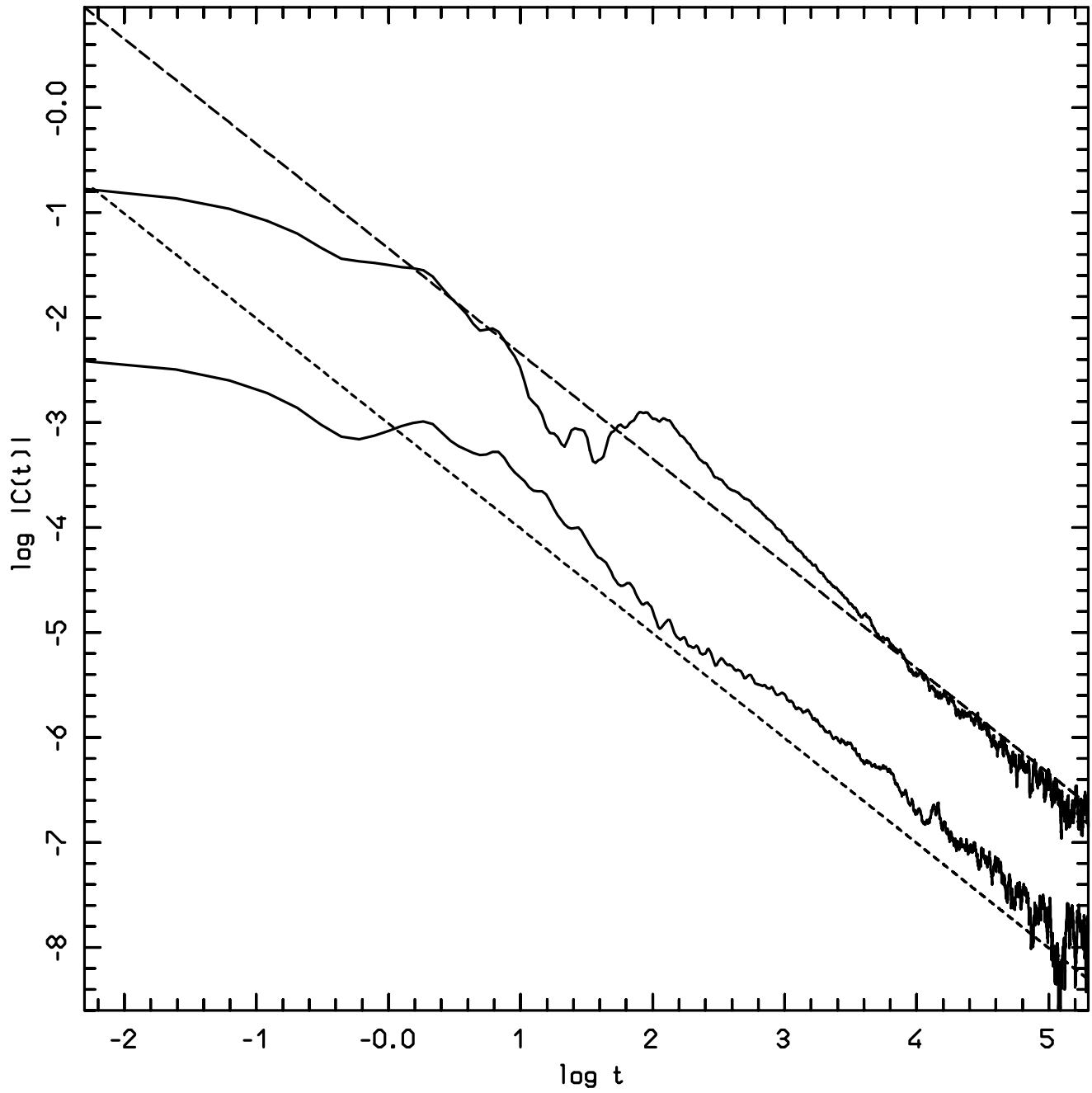
\end{document}